\documentclass[letterpaper,twocolumn,aps,prl,showpacs,amsmath,amssymb,floatfix]{revtex4-1}
\usepackage{graphicx}
\usepackage{dcolumn}
\usepackage{bm}

\begin{document}

\title{The Geometry of the Vapor Layer Under a Leidenfrost Drop}

\author{J.C. Burton}
\email{jcburton@uchicago.edu}
\author{A.L. Sharpe}
\author{R.C.A. van der Veen}
\author{A. Franco}
\author{S.R. Nagel}

\affiliation{James Franck Institute and Department of Physics, The University of Chicago}

\date{\today}

\begin{abstract} 
In the Leidenfrost effect, liquid drops deposited on a hot surface levitate on a thin vapor cushion fed by evaporation of the liquid.  This vapor layer forms a concave depression in the drop interface.  Using laser-light interference coupled to high-speed imaging, we measured the radius, curvature, and height of the vapor pocket, as well as non-axisymmetric fluctuations of the interface for water drops at different temperatures.  The geometry of the vapor pocket depends primarily on the drop size and not on the substrate temperature.
\end{abstract}


\maketitle

\begin{figure}[!]
\begin{center}
\includegraphics[width=7.0 cm]{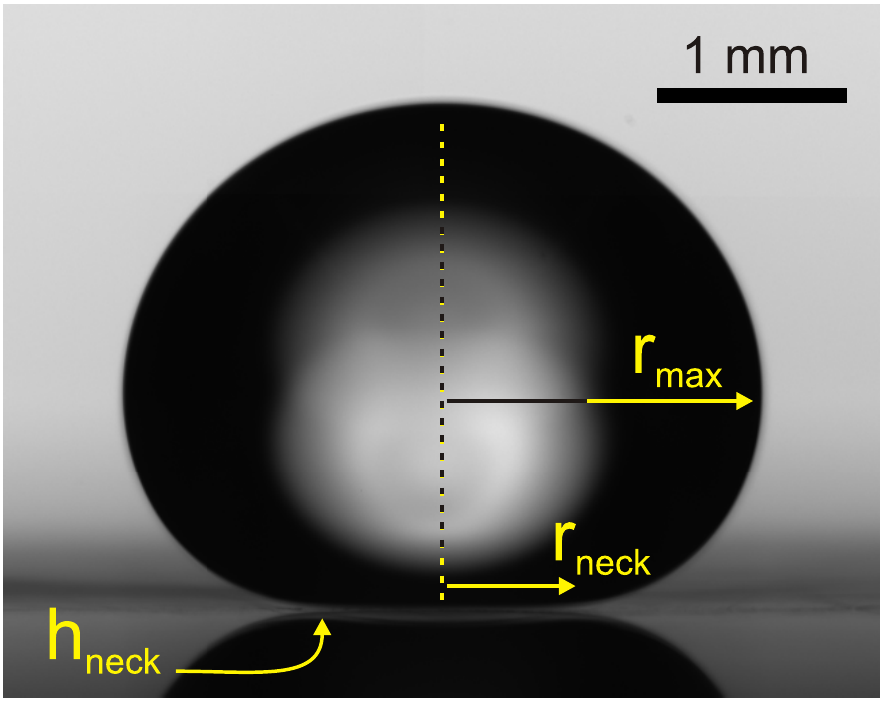}
\caption[]{Image of a water Leidenfrost drop from the side at substrate temperature $T_s$ = 320$^\circ$C. The arrows indicate the maximum drop radius $r_{max}$, the radius of the neck $r_{neck}$, and the height of the neck from the surface $h_{neck}$.
} 
\label{sideimg_fig}
\end{center} 
\end{figure}

The nearly frictionless motion of  water droplets on a hot pan is familiar to any cook. Such Leidenfrost drops \cite{Leidenfrost1756} levitate on a thin, insulating cushion of evaporated vapor (Fig.\ \ref{sideimg_fig}). Above a Leidenfrost temperature $T_L$ there is a sudden increase in the drop lifetime \cite{Gottfried1966,Bernardin1999,Biance2003} and the cessation of nucleated boiling. The stable vapor film, fed by the drop evaporation, drains under a thin annulus that resides closest to the substrate. Large Leidenfrost drops can undergo ``star"-shaped oscillations \cite{Brunet2011,Tokugawa1994,Adachi1984,Group2000}, and a Leidenfrost vapor layer can reduce the drag on a hot solid falling in a fluid \cite{Vakarelski2011}. In addition, Leidenfrost drops can propel themselves upon a ratcheted surface \cite{Linke2006,Lagubeau2011,Dupeux2011,Cousins2012} as well as deposit micron-sized trails of nanoparticles \cite{Elbahri2007}. Using light interference coupled to high-speed imaging, we measure the geometry of the vapor layer under a Leidenfrost drop. Although there are large fluctuations in its shape, over a wide range of temperatures a sustained vapor pocket exists whose average properties do not depend strongly on the evaporative flux from the drop.

\begin{figure}[!]
\begin{center}
\includegraphics[width=8.6 cm]{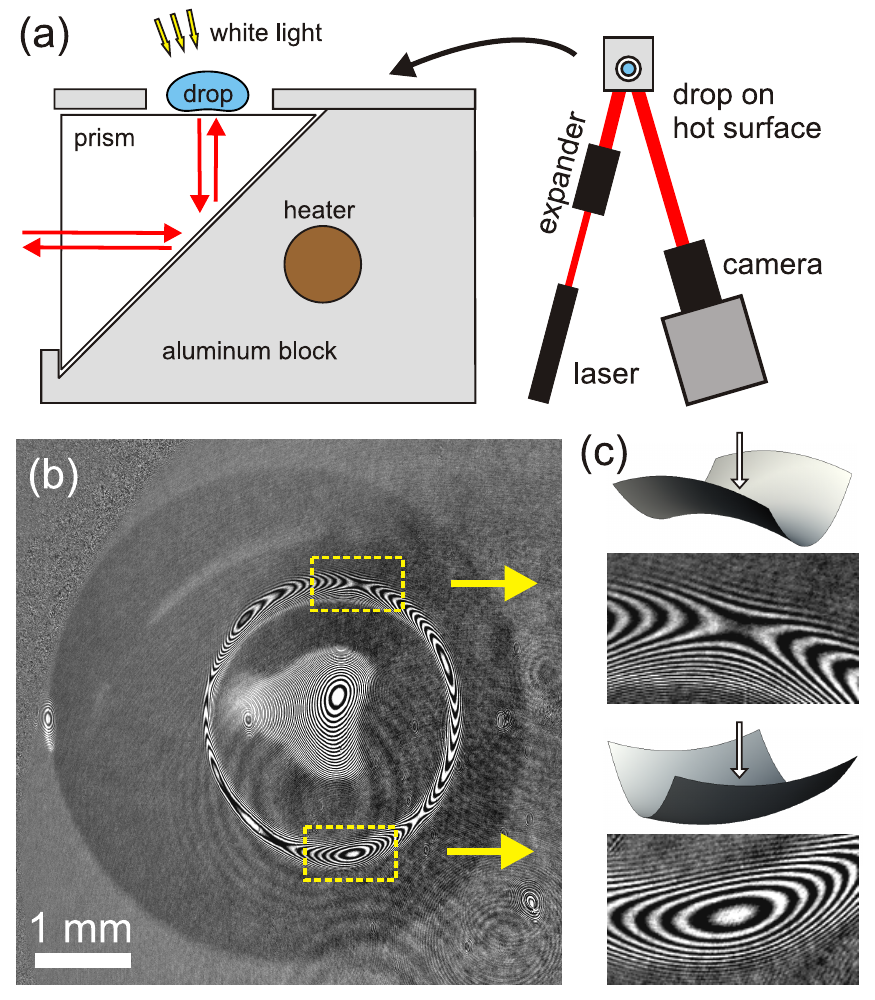}
\caption[]{(a) Schematic of measurement. Drops are deposited on a prism surface encased in a heated aluminum block. Reflected light from a He-Ne laser 
causes interference between the bottom surface of the drop and the top surface of the prism. White-light illumination enters the prism at a slight angle to produce a shadow that is slightly offset from the interference pattern. (b) Image from beneath a water Leidenfrost drop at $T_s$ = 245$^\circ$C. The dark circular region is the shadow produced by the white light. Fringes are most visible in the neck region (outer ring interference pattern) and at the center of the gas pocket (inner concentric fringes). (c) Saddle interference patterns around the neck are due to surfaces with negative Gaussian curvature; circular patterns are due to surfaces with positive Gaussian curvature.
} 
\label{expsetup_fig}
\end{center} 
\end{figure}

\begin{figure}[!]
\begin{center}
\includegraphics[width=8.6 cm]{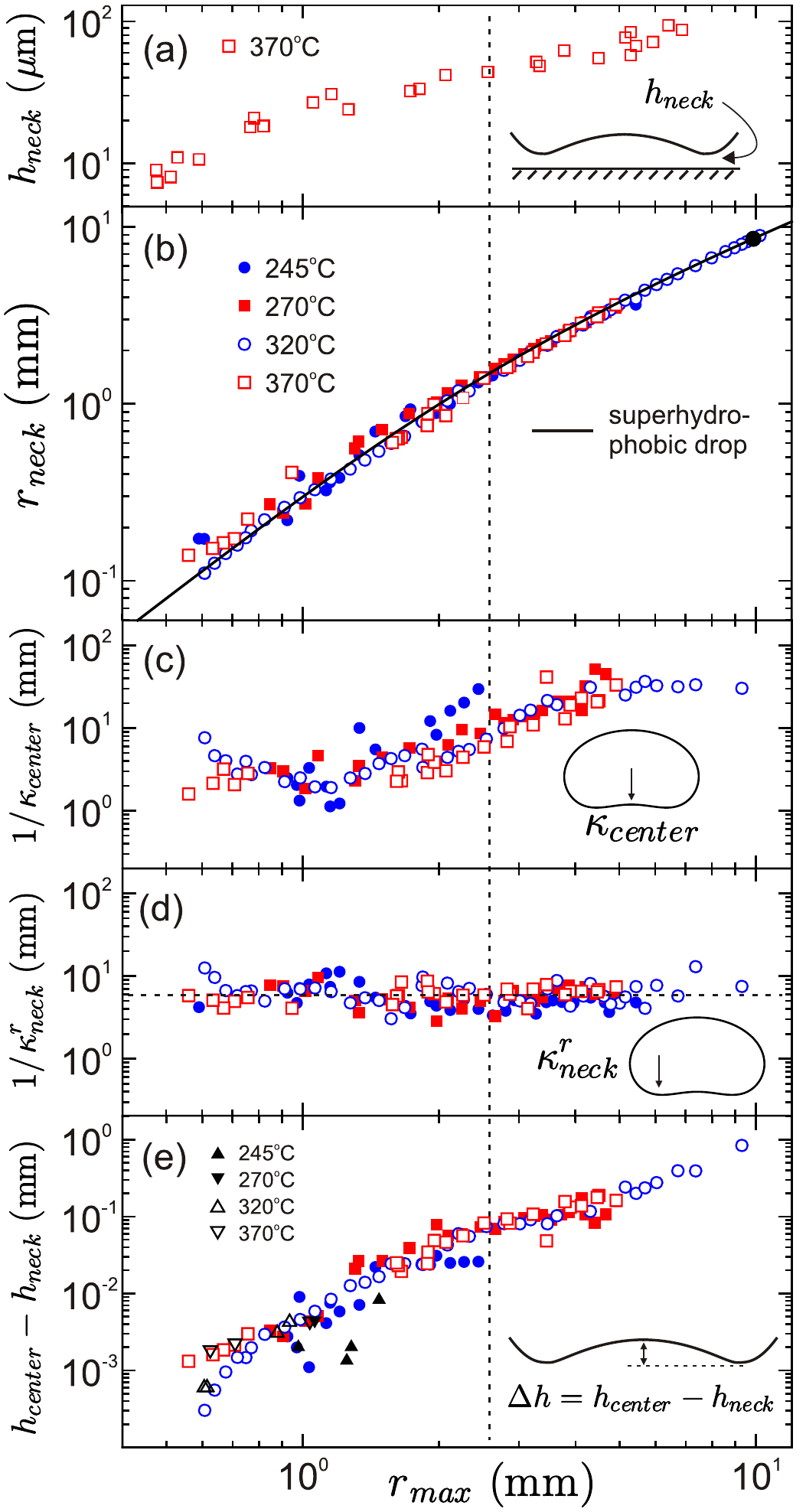}
\caption[]{Measurements of the vapor layer versus $r_{max}$.
(a) Height of the neck measured from horizontal images.  (b) Neck radius. The black line shows $r_{max}$ from a model of a superhydrophobic drop. The solid black dot is the predicted maximum drop size \cite{Snoeijer2009}.
(c) Average radius of curvature of the center and (d) radial radius of curvature of the neck. The horizontal dotted line in (d) represents 2.3$\lambda_c$. (e) Thickness of the vapor pocket. All triangles are measured from drops with a continuous interference pattern from the neck to the center. Circles and squares are calculated using eqn.\ \ref{heightdiff_eq}. The dashed vertical line represents $r_{max}=\lambda_c=2.5$ mm. 
} 
\label{dropdata_fig}
\end{center} 
\end{figure}

We used reflected, monochromatic light to image thin vapor layers at high speeds, a technique we previously developed to study air layers beneath splashing droplets \cite{Driscoll2011}. In the setup shown in Fig.\ \ref{expsetup_fig}a, a fused-silica prism is encased in an aluminum block covering all its sides except the front face and a circular hole on the top. The temperature of the prism was regulated up to $T_s$ = 370$^\circ$C. Above the Leidenfrost temperature $T_L\approx160^\circ$C, drops of deionized water were deposited and were free to move in the circular hole. They were imaged until they completely evaporated, which occurred after a few minutes. The average temperature in the levitated drop was always 99 $\pm$ 0.5$^\circ$C as measured with a thermistor. Near boiling, the density of water is $\rho$ = 958 kg m$^{-3}$ and the surface tension is $\gamma$ = 59 mN m$^{-1}$, so that the capillary length $\lambda_c=\sqrt{\gamma/\rho g}$ = 2.5 mm. For typical drop sizes used in our experiments, the gap height $h_{neck}$ between the hot substrate and the bottom of the drop ranged from 5 $\mu$m to 100 $\mu$m (Fig.\ \ref{dropdata_fig}a). 

We use a He-Ne laser with wavelength $\lambda_l$ = 632 nm and coherence length $>$ 1 m.  The beam was expanded to a diameter of $\sim$ 3 cm and sent into a prism at a small angle to prevent extraneous interference patterns generated between the faces of the prism and other optical elements. Only the gap between the substrate and the drop surface produced fringes. Approximately 3.5\% of the light is reflected from the top surface of the prism and 2.1\% from the bottom surface of the drop. The reflected light is imaged by a high-resolution camera (Prosilica GX3300) with an exposure time less than 100 $\mu$s. High resolution was necessary to minimize Moir\'{e} effects where the fringe spacing is comparable to the pixel size. Simultaneously, we measure the drop radius, $r_{max}$, with white-light illumination from above and incident at a slight angle so that it reflects into the camera. (White-light interference, used to measure the very thin vapor layers in drop impact~\cite{Tran2012}, cannot be used here because static Leidenfrost vapor layers are much thicker.)  
A similar experimental setup was recently used to study the collapse of the Leidenfrost vapor layer under an applied electric field \cite{Celestini2012}.

The reflected light from the glass and liquid surfaces can interfere either constructively or destructively; the intensity is approximately $I\propto\sin\left(2\pi h/\lambda_l\right)^2$, where $h$ is the distance between the substrate and the liquid surface. Thus the height difference between a black and a white fringe is $\lambda_l/4$ = 158 nm.  As shown in Fig.\ \ref{expsetup_fig}b, fringes will be most visible in regions that are nearly flat (\textit{i.e.}, $\nabla h$ $\approx$ 0). This occurs near the center of the gas pocket and in the thin annular neck surrounding it. This method can measure only variations in $h$ but not its absolute value. Because the gas pocket acts as a weakly-focusing convex mirror, the center is surrounded by a bright region. 

The fringes encode the geometry of the liquid surface. Fig.\ \ref{expsetup_fig}b-c shows the fringe pattern created by curvature around the neck of the drop. Saddle patterns are produced when the Gaussian curvature is negative, and circular patterns are produced when the Gaussian curvature is positive. Since the neck is closest to the surface, the curvature in the radial direction must be positive; thus the circular patterns around the neck represent local minima. 

Fig.\ \ref{dropdata_fig}a shows measured values of $h_{neck}$ from horizontal images, such as Fig.\ \ref{sideimg_fig}. (This technique only works for high $T_s$ where $h_{neck}/r_{neck}$ is large so that rear illumination can pass under the drop without significant scattering.) Fig.\ \ref{dropdata_fig}b shows the size of the neck at 4 different surface temperatures, $T_s$  (and thus different evaporative gas fluxes). The neck radius, $r_{neck}$, is nearly independent of $T_s$ and depends primarily on the drop radius $r_{max}$. There is significant scatter in the data for small drop sizes. At all temperatures, small drops sometimes bounce with an amplitude that is a significant fraction of the drop radius. We are unaware of any reported studies of this phenomenon. However, we speculate that the bouncing is due to coupling between the evaporation and capillary oscillations of the drop.

We estimate the shape of a stable, isothermal Leidenfrost drop in the limit of negligible internal flow following ref.\ \cite{Aussillous2001,Biance2003}. Most of the drop is surrounded by a constant atmospheric pressure. Therefore, $r_{max}$ and $r_{neck}$ depend on gravity and surface tension, identical to that of a drop on a superhydrophobic surface, where the liquid interface is tangent to the substrate at $r_{neck}$. Solving the Young-Laplace equation to obtain drop shapes \cite{Burton2010}, we obtain for the superhydrophobic model the solid line in  Fig.\ \ref{dropdata_fig}b which is in excellent agreement with the data.

\begin{figure*}[!]
\begin{center}
\includegraphics[width=16.5 cm]{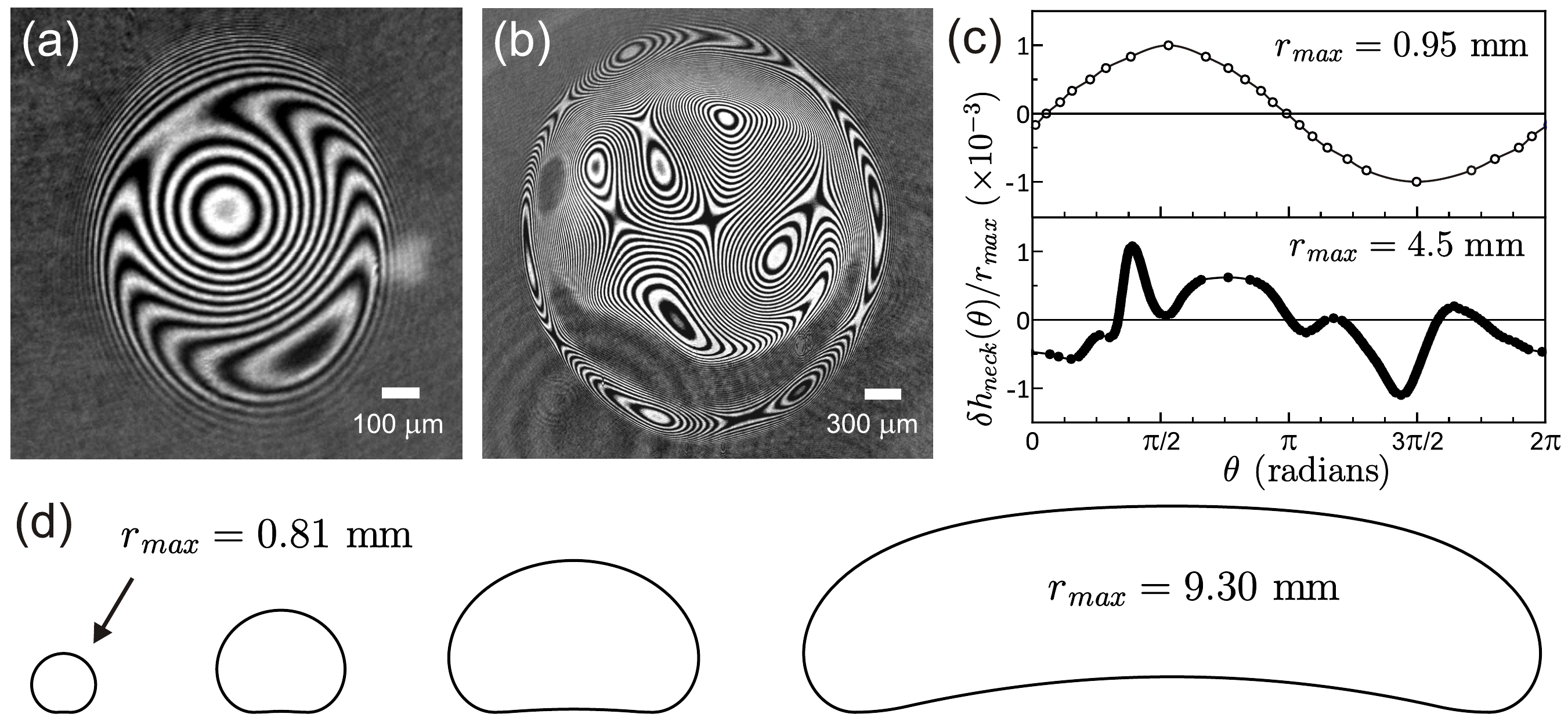}
\caption[]{(a-b) Interference patterns at a surface temperature $T_s$ = 245$^\circ$C of a small water drop with $r_{max}$ = 1.2 mm (a), and a larger water drop with $r_{max}$ = 2.7 mm (b). The larger drop shows fluctuations in the inner gas pocket leading to a complex interference pattern. (c) Variation of the neck height around its circumference for drops with $T_s$ = 370$^\circ$C. Small drops with $r_{max}\ll\lambda_c$ have one wavelength of amplitude variation. Larger drops have higher modes of amplitude variation. (d) Reconstruction of axisymmetric Leidenfrost water drops of various sizes. 
} 
\label{dropgeometry_fig}
\end{center} 
\end{figure*}

Large drops have a thickness of order $2\lambda_c$ \cite{Biance2003}.  There is a maximum size for Leidenfrost and gas-levitated drops before the gas pocket bubbles upward and turns the drop into a torus \cite{Biance2003,Snoeijer2009,Duchemin2005}. When this occurs, the gas pocket repeatedly rises and bursts through the top of the drop with a period of $\approx 1$ sec, which is discussed in detail in ref.\ \cite{Biance2003}. The maximum neck radius of a Leidenfrost drop is predicted to be $r_{neck}=r_{max}-0.53\lambda_c$ when $r_{max}\approx3.95 \lambda_c$, the maximum size of the drop at the initiation of the instability \cite{Snoeijer2009}.  Our measurements of the maximum drop size and neck radius in Fig.\ \ref{dropdata_fig}b are within 4\% of this prediction.  

For small drops, $r_{neck}\approx 0.8 r_{max}^2/\lambda_c$, in agreement with \cite{Aussillous2001,Biance2003}. To calculate this prefactor, we consider a small, axisymmetric Leidenfrost drop with $r_{max}\ll\lambda_c$. In this limit gravity has a negligible effect so that the Laplace pressure inside the drop is: $P_{in}=2\gamma/r_{max}$. Underneath the drop, the pressure is $P_{out}=P_{in}-2\gamma\kappa_m$, where $\kappa_m$ is the mean curvature of the drop/vapor-layer interface. If $\kappa_m\ll 1/r_{max}$, then $P_{out}\approx P_{in}$. As we show below, this is the case for small drops. This pressure must support the weight of the drop over the neck region: $\frac{4}{3}\pi r_{max}^3 \rho g=\pi r_{neck}^2 P_{in}$.  Thus for asymptotically small drops we have the exact result
\begin{align} 
r_{neck}=\sqrt{\frac{2}{3}}r_{max}^2/\lambda_c\approx0.82r_{max}^2/\lambda_c.
\label{rnecksmall_eq} 
\end{align}


The curvature of the drop interface can be measured in regions where fringes are visible. Fig.\ \ref{dropdata_fig}c shows the curvature at the center of the gas pocket, $\kappa_{center}$, averaged over two perpendicular directions given by the major and minor axes of the elliptical fringe pattern. Fig.\ \ref{dropdata_fig}d shows the curvature of neck in the radial direction, $\kappa^r_{neck}$, averaged over 4 points around the circumference of the neck. We find that $1/\kappa^r_{neck}\approx2.3\lambda_c$ over our entire range of data, while $1/\kappa_{center}$ seems constant at both small and large drop sizes, but changes by a factor of $\sim$ 10 between these regions. Both radii of curvature appear independent of $T_s$ and are significantly larger than $r_{max}$ for small drops, so that eqn.\ \ref{rnecksmall_eq} is valid. 

For small drops we can observe the entire interference pattern from the neck to the center (Fig.\ \ref{dropgeometry_fig}a). By counting fringes we measure the difference in height, $\Delta h \equiv h_{center}-h_{neck}$, which is indicated in Fig.\ \ref{dropdata_fig}e by solid and open black triangles. For larger drops we can estimate  $\Delta h$  by assuming a simple model for the geometry of the vapor pocket. Given two parabolas centered at $r=0$ and $r=r_{neck}$ with respective curvatures $-\kappa_{center}$ and $\kappa^r_{neck}$, a unique value of $\Delta h$ joins the two parabolas with a continuous derivative:
\begin{align} 
\Delta h = \frac{r_{neck}^2\kappa_{center}\kappa^r_{neck}}{2(\kappa_{center}+\kappa^r_{neck})}.
\label{heightdiff_eq} 
\end{align} 

The squares and circles in Fig.\ \ref{dropdata_fig}e are points calculated using eqn.\ \ref{heightdiff_eq} and the measured values of $\kappa_{center}$, $\kappa_{neck}^r$, and $r_{neck}$. We find that $\Delta h$ varies by over 3 decades. 
Although we observe that $h_{neck}$ depends on the gas flux and thus the substrate temperature, $T_s$, the data from Fig.\ \ref{dropdata_fig} show that the geometry of the vapor pocket is nearly independent of $T_s$.

We next examine deviations from the axisymmetric description of a Leidenfrost drop.  At higher $T_s$ and smaller $r_{max}$, there is only one extrema associated with the top of the vapor pocket. Fig.\ \ref{dropgeometry_fig}a shows an interference pattern for a small drop $r_{max}$ = 1.2 mm, with a concave center region surrounded by the annular neck.  However, at lower $T_s$ and larger $r_{max}$, there are multiple extrema indicating a complex interfacial geometry as seen in Fig.\ \ref{dropgeometry_fig}b. 
We interpret this as increasing pressure fluctuations when $h_{neck}$ is small (low $T_s$),  or when $r_{neck}\gg\lambda_c$ so that higher-order perturbations are no longer strongly suppressed by surface tension.   

Fig.\ \ref{expsetup_fig}b and Fig.\ \ref{dropgeometry_fig}a-b show fringes around the circumference of the neck that allow us to calculate the height variation of the surface $\delta h_{neck}(\theta)$ versus azimuthal angle $\theta$. Fig.\ \ref{dropgeometry_fig}c shows $\delta h_{neck}(\theta)/r_{max}$ for drops of different sizes. For small drops, there is a single wavelength variation around the neck. For larger drops, higher-order modes produce more complex profiles. Using data for the average value of $h_{neck}$ from Fig.\ \ref{dropdata_fig}a, we estimate that $\delta h_{neck}(\theta)$ can be as large as 20\% of the average value of $h_{neck}(\theta)$. 

In conclusion, our measurements allow us to produce a {\it quantitative} picture of the Leidenfrost drop geometry. Using the curvature data from Fig.\ \ref{dropdata_fig}, we can combine the two smoothly-joined parabolas (eqn.\ \ref{heightdiff_eq}) with the shape of a superhydrophobic drop at $r=r_{neck}$. Four different drop sizes are shown in Fig.\ \ref{dropgeometry_fig}d, showing spherical drops with flat bottoms at small $r_{max}$, and drops near the onset of the instability at large $r_{max}$. 

Counterintuitively, a drop on a very hot surface will survive for a surprisingly long time because it is levitated on an insulating layer of vapor. We have shown that the geometry of this insulating layer is nearly independent of substrate temperature, and depends only on the drop size. This vapor layer becomes unstable for large drops.
Moreover, significant fluctuations exist in the central vapor pocket and around the neck region by which the vapor escapes. These fluctuations vary quickly in time and space, and may be important for drop propulsion, oscillations and the ``star"-shapes observed at high temperature.

We thank Michelle Driscoll, Taehun Lee and Jeff Morris for valuable discussions. We acknowledge support from NSF-PREM DMR-0934192 (JCB), NSF-MRSEC DMR-0820054 (RCAV), NSF DMR-1105145 (SRN), NSF REU PHY-1062785 (ALS), and NSF Materials World Network DMR-0807012 (AF).

\end{document}